\documentclass[final,5p,times,twocolumn,sort&compress]{elsarticle}

\usepackage{amssymb}
\usepackage{amsmath}
\usepackage{lipsum}
\usepackage{graphicx}
\usepackage{caption}
\usepackage{subcaption}
\usepackage{epstopdf}
\usepackage{xcolor}  
\usepackage[normalem]{ulem}
\usepackage[subrefformat=parens]{subcaption} 
\journal{Physics Letters B}

\begin{document}
	
\begin{frontmatter}	
		
\title{Impact of microscopic structural transitions on particle stability and lifetimes of hot nuclei}	 

\author[1]{Mamta Aggarwal\corref{cor1}}
\ead{mamta.a4@gmail.com}
\cortext[cor1]{Corresponding author}

\author[1]{Pranali Parab}
\author[2]{G. Saxena\corref{cor2}}
\ead{gauravphy@gmail.com}
\cortext[cor2]{Co-corresponding author}

\address[1]{Department of Physics, University of Mumbai,
             Vidya Nagari, Mumbai 400098, Maharashtra, India.}
\address[2]{Department of Physics (H \& S), Govt. Women Engineering College, Ajmer 305002, Rajasthan, India.}

\begin{abstract}
The impact of temperature-induced deformations and shape fluctuations on the particle stability and decay processes has been investigated across the isotopes of hot nuclear systems with $Z = 28$ to $50$, with focus on astrophysically crucial pathways at excitation energies relevant to stellar environments. We perform global finite-temperature analysis using the statistical theory of hot nuclei combined with the triaxially deformed Nilsson Hamiltonian and Strutinsky's prescription, and explore the interplay between deformation, shell quenching, separation energies, and $\beta$-decay characteristics at finite temperatures. Our results show that around critical temperatures $T_c \approx 1$--$2$ MeV, where the shell quenching effects become predominant, the nuclear deformation reduces and the shape undergoes a transition to the spherical configuration. Our computed neutron and proton separation energies, which usually decrease with increasing temperature, implying the reduced binding in hot nuclei, occasionally show an enhancement in some nuclei at reduced deformation around $T_c$ that shifts the last unbound nucleon to the bound, stabilizing the nucleus by shifting the drip-line boundaries. A few nuclei are found to show one- and two-neutron drip line expansion with temperature. Moreover, the temperature-induced changes in deformation strongly correlate with the marked variations in our calculated $Q_\beta$ values and lifetimes, underscoring their impact on weak interaction rates. These findings provide insight into the sensitivity of particle stability and weak-interaction observables to thermal effects and may serve as complementary inputs for modeling nuclear processes in hot astrophysical environments. 
\end{abstract}

\begin{keyword}
Deformation \sep Separation Energy \sep Shell Effects \sep Q-values \sep $\beta$-decay Lifetimes \sep Nucleosynthesis
\end{keyword}
		
\end{frontmatter}
	
\section{Introduction}
\label{introduction}
Nuclear processes are known to generate energy and drive astrophysical phenomena of nucleosynthesis \cite{ShenPPN21} in high temperature environments through various thermonuclear reactions ~\cite{ThielmannPPN07}. Since the beginning of the universe, the onset of the element formation has been defined by the nuclear masses and microscopic structure ~\cite{B2FH} with protons ($^1H$), neutrons,  and Helium ($^4He$) as the fundamental nuclear building blocks for matter. The inputs from  microscopic nuclear structure, deformation, and decay properties play a significant role in determining the timescales of stellar processes ~\cite{B2FH,Cowan,Sasaki} that may range from stellar evolution over millions of years to comparatively rapid stellar explosions. This is a unique kind of an example of the interplay between the microscopic nuclear behaviour and macroscopic celestial events. In the extreme scenarios of supernovae or neutron star mergers where the temperatures may be as high as 2 MeV \cite{JankaPR06}, the structural properties of hot nuclei influence the nucleosynthesis dynamics profoundly. The hot nuclei ~\cite{MRSPRC88,GoodmanNPA91,AlhassidPRL86} that are studied as the excited compound nuclear systems ~\cite{RauscherPRC10}, are characterized by the excitation energy and temperature. The de-excitation of such hot nuclear systems via particle evaporation or other decay channels are influenced by the structural dynamics  ~\cite{MAPLB10,MAPRC14,MANPA24} with their shorter lifetimes that pose the obvious challenges in their study. This calls for reliable nuclear theoretical models in close connection with the experimental nuclear physics and astrophysics.\par

The probes to understand the dependency of nuclear structural effects on nucleosynthesis processes is currently one of the most thrust areas of nuclear astrophysics ~\cite{YuPRL24, Muli25}. The difference of masses of the parent and residual nuclei determine the energy generation and the possible decay channels that may be energetically allowed via nucleon evaporation, $\alpha$-~\cite{GamboaPRC24},  $\beta$-~\cite{SuzukiPRC12}, $\gamma$- or heavier cluster decay~\cite{WardaPRC18}. Heavy-element abundances~\cite{HolmbeckEPJA23} are created by neutron-capture processes competing with $\beta$ decay. In s-process, where the neutron-capture timescale is much longer than the $\beta$-decay lifetime, nucleosynthesis proceeds along the $\beta$-stable nuclei~\cite{KappelerRMP11}, whereas the r-process with neutron capture time $<<$ $\beta$-decay lifetime produces unstable nuclei toward the neutron drip line~\cite{QianNPA04}. The rp-process produces neutron-deficient unstable nuclei approaching the proton drip line~\cite{HoveFBS17,ZhouN23, SchatzPR98}. The well known s- and r- process peaks in neutron-rich isotopes at magic numbers N = 50, 82, 126~\cite{MayerARAA94} observed in the element abundance distribution curves reflect the significant impact of shell effects~\cite{ZhiPRC13}  that start diminishing at finite temperatures and the nuclear properties begin to resemble that of a liquid drop or an ideal Fermi gas~\cite{MengoniJNST94}. In high temperature environments relevant to stellar events, understanding of such nuclear processes requires a systematic theoretical treatment of structural aspects of excited nuclear states  ~\cite{SchatzJPG22} which is the objective of present study. \par

The nuclear masses and binding energies along the stability valley are known to us to an extent~\cite{MollerANDT19} due to various advances in experimental techniques that have enabled crucial inputs on  nuclear masses, lifetimes ~\cite{HorowitzJPG19, GalesJP11,ThoennesseRPP13}, thermodynamical nuclear properties and shapes ~\cite{BackRMP14,SnoverARNPS86}, but the understanding of  weakly bound or unstable nuclei close to drip lines is still a challenge in the experimental nuclear astrophysics~\cite{ThoennessenRPP04} and one needs to rely on predictions of theoretical models that can properly treat the interplay of nuclear bulk properties and quantum shell effects at finite temperatures ~\cite{MAPLB10}. Several microscopic and macroscopic approaches are popularly known to study temperature-dependent nuclear properties~\cite{GoodmanNPA81, EdigoPLB97, AgrawalNPA03, NakadaPRC11,KanekoPRC06,RavlicPRC24, ChenChinP23, WangPRC21}. However, the statistical descriptions of hot nuclei~\cite{MRSPRC88, KatariaPRC78, KatariaNPA80, MRSP89}, including our applications in earlier works~\cite{MAPLB10, MAPRC14, MAPRC98} has been quite successful in incorporating the structural effects in compound nuclear properties which demonstrated that the nuclear deformation, shapes, and binding energies appear to be sensitive to thermal effects, which makes it suitable to explore the unknown domains of nuclear processes relevant in astrophysics. \par

This article delves into the temperature-driven shape dynamics of isotopic chains of nuclei $Z = 28$--$50$ ranging from proton to neutron drip line, combined with their unique nuclear shapes ranging from spherical to deformed prolate, oblate and triaxial shapes~\cite{LeoniEPJ24} and shape phase transitions~\cite{EsmaylzadehPRC19}, which offers valuable clues about the conditions under which these isotopes decay or stabilize. Nuclear deformation and shape are a crucial nuclear structure input in astrophysical modelling as they influence the prediction of masses~\cite{MAPRC14}, Q-values~\cite{MANPA24,GSJPG23, GSSR24}, level  densities~\cite{MANPA19-23}, nucleon emission probability~\cite{SguazzinPRL25}, and $\beta$-decay half lives~\cite{MANPA24}. Nuclear isotopic chain of $Z=28$ to 50, synthesized in various nucleosynthesis processes, presents a diverse range of nuclei with magic, semi magic, doubly magic and deformed nuclei. Well deformed nuclei across the region $Z = 28$--$50$, in particular, near mass region A $\approx$ 80$-$120~\cite{Hua,Sarriguren,Abusara} are known to exhibit a variety of deformed states with rapid shape phase transitions and shape coexistence ~\cite{LalazissisNPA95, RodriguezPLB10} and hence provide an ideal laboratory to study the role of nuclear deformation and shape effects~\cite{SkalskiNPA97,KirchukPRC93} in nuclear stability and decay processes.\par

Our recent works have shown the impact of deformation and shape coexistence on particle stability for 1p-decay half lives~\cite{GSSR24} for nuclei with Z $<$ 82,  and $\beta$-decay half lives~\cite{MANPA24} for Mo-Ru isotopes that demonstrated the role of nuclear structural effects on lifetimes in ground state nuclei. Incorporating the temperature effects ~\cite{RoutPRL13, WibowoPRL22, ParmarPRC22, RenPLB24} induces a variety of deformation and shape fluctuations~\cite{MAPRC14,MAPRC98,MTN, MNV} which impact the nuclear binding that may alter the precise position of drip lines as indicated in our earlier works ~\cite{MAPLB10,MAPRC98}. A recent work has reported the drip line expansion with increasing temperature ~\cite{RavlicNC23} while citing the shell quenching effects as the possible reason, but a clear link between the effects due to shell quenching and the location of drip lines is not shown. To get better insights on this subject, we investigate the structural effects in hot nuclei especially close to drip lines and explore a correlation between the deformation and shapes of parent and daughter nuclei ~\cite{GSSR24} on particle stability, binding, Q-values and lifetimes of hot nuclei. The Q-values associated with $\beta^{+}$ and $\beta^{-}$ - decay and their lifetimes are found to be sensitive to the shift in deformation at high temperatures that impact decay energetics. In nuclei far from stability, where the uncertainties are larger, these effects are further enhanced. A shift in the one- and two- neutron drip lines with
temperature induced structural changes is observed. By relating the deformation and finite-temperature effects to Q-values, $\beta$-decay lifetimes and precise position of drip lines, we aim to refine the nuclear inputs for astrophysical reactions and our understanding of how microscopic nuclear features influence the lifetimes and dynamics of astrophysical processes.\par

\section{Theoretical Approach}

We perform the global finite-temperature analysis across the region bounded by the two major shell closures $Z = 28$ and $50$ and study the interplay between shell effects, deformation and temperature effects in an excited compound nuclear system in thermal equilibrium for isotopic chains 	$^{48-108}\mathrm{Ni},\ ^{54-112}\mathrm{Zn},\ ^{60-114}\mathrm{Ge},\ ^{64-116}\mathrm{Se},\ ^{68-120}\mathrm{Kr},\ ^{72-128}\mathrm{Sr},\allowbreak\ ^{76-138}\mathrm{Zr},\ ^{78-144}\mathrm{Mo},\ ^{84-150}\mathrm{Ru},\ ^{86-158}\mathrm{Pd},\ ^{92-168}\mathrm{Cd},\ ^{98-174}\mathrm{Sn}$ over a wide range of temperature $T = 0.6 - 3.0$ MeV which is suitable to understand the high temperature environments in stellar processes~\cite{NishimuraMNRA19}. We employ Microscopic Statistical Model combined with Macroscopic-Microscopic approach using the triaxially deformed Nilsson-Strutinsky Model (NSM) ~\cite{MAPLB10,MAPRC14} where the excited compound nuclei are described as the thermodynamical system of fermions incorporating the deformation, shape and temperature degrees of freedom.The statistical temperature ($T$) is taken as input parameter. To determine the nuclear deformation and shape of the hot nuclei, it is common to minimize the appropriate free energy F (F $=$ E $-$ TS)~\cite{GoodmanPRC88}. We trace F minima with respect to the Nilsson deformation parameters $\beta_2$ and $\gamma$ which give the deformation and shape of excited nuclei at finite temperatures~\cite{GoodmanPRC88}. Energy (E), entropy (S) and the excitation energy $(E^*)$, are the functions of particle number, deformation and shape, and are derived from the statistical properties of single-particle levels at finite temperature (see Eqs. (1) $-$ (10)) of theoretical formalism in supplementary Note 1). The calculations are performed for the angular deformation parameter $\gamma$ ranging from $-180^{\circ}$ (oblate) to $-120^{\circ}$ (prolate), including triaxial shapes in the interval $-180^{\circ} < \gamma < -120^{\circ}$. The axial deformation parameter $\beta_2$ is varied from 0 to 0.4 in steps of 0.01. (Details of this formalism adequately described in our earlier works~\cite{MAPLB10,MAPRC14, MAPRC10} have been provided in supplementary Note 1 for the reader's quick reference).

\section{Results and Discussion}

\begin{figure}[htbp]
	\centering 
	\includegraphics[width = 0.495\textwidth]{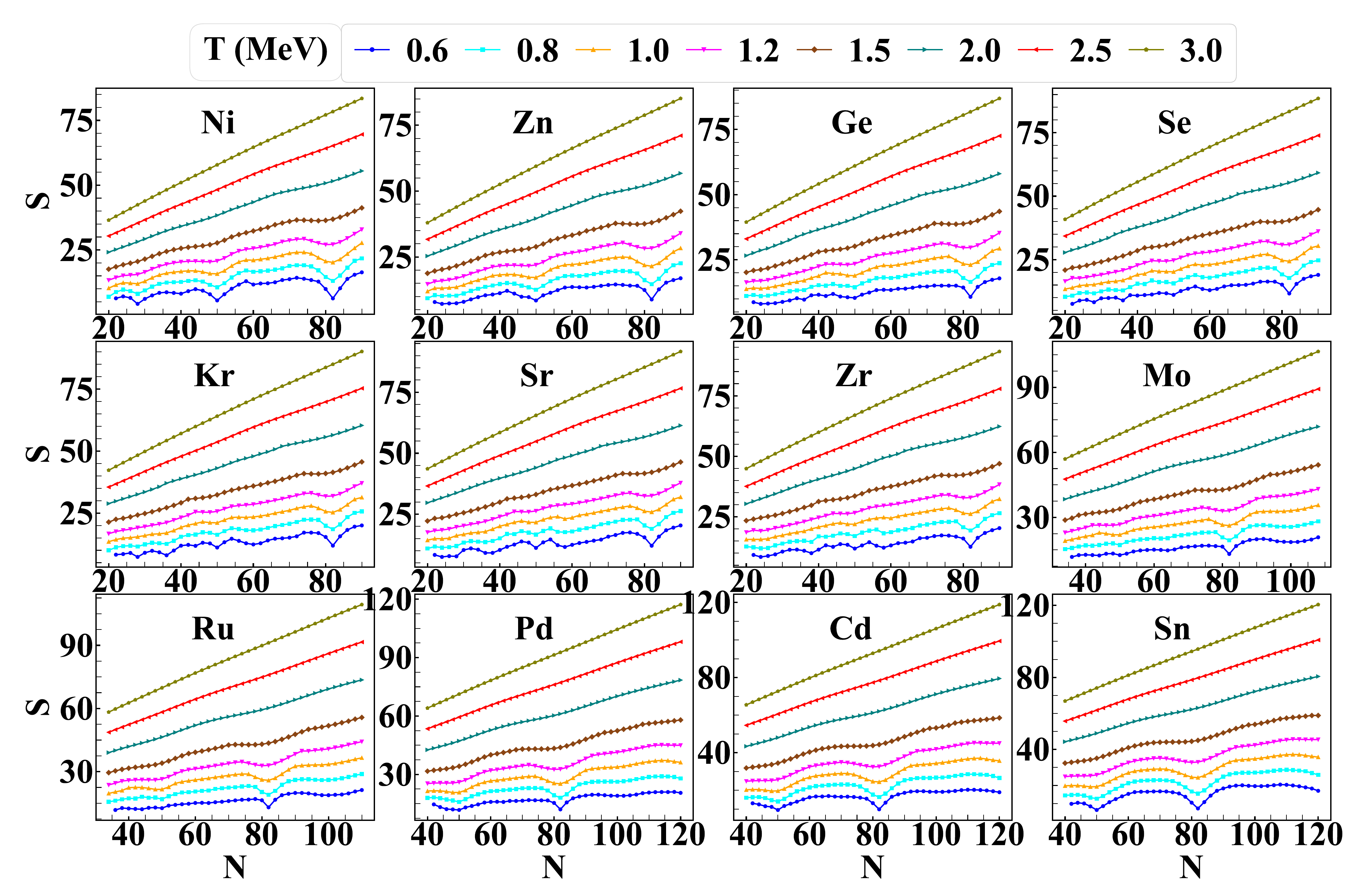}	
	\caption{Variation of entropy (S) with temperature for isotopic chains ranging from Z = 28 to 50} 
	\label{figEntropy}%
\end{figure}

As we know that in case of a hot nucleus, there is a threshold to excitation energy E$^*$ below which a nucleus can remain a bound system~\cite{BoncheNPA85} and one can probe the nuclear phenomena like giant pairing vibrations~\cite{BarrancoPRL25} or the disappearance of shell effects in hot nucleus \cite{MAPRC98}. With increasing temperature, the internal excitation of nucleons increases, shell effects weaken, pairing correlations break down, reducing the order, enabling the nucleons to populate a broader range of states, which results in increasing the nuclear entropy as seen in Fig. \ref{figEntropy}, which shows variation of entropy as a function of neutron number and T for the isotopes of Ni, Zn, Ge, Se, Kr, Sr, Zr, Mo, Ru Pd, Cd and Sn.  Entropy, an important thermodynamical parameter that measures the number of accessible microstates available to a hot nuclear system, increases roughly linearly with the temperature except for fluctuations at closed shells (N $=$ 28, 50, 82) where it takes more energy to excite nucleons across the larger shell gaps. This effect is particularly pronounced at low temperatures which gets smeared out with increasing temperature due to disappearing shell effects and pairing correlations. Although the pairing correlations influence the structural properties to an extent, but since the temperature range explored here lies predominantly in the regime (T $=$ 0.6 $-$ 3.0 MeV) where the pairing correlations are expected to be strongly suppressed or become negligible, the pairing effects are not incorporated in this study. As demonstrated in earlier studies~\cite{RavlicPRC21, LisboaPRC16, NiuPRC13}, the pairing effects fade away quite rapidly with increasing thermal energy, whereas the shell and deformation effects continue to govern the structural evolution of hot nuclei, we focus on shell effects in this work, in particular, the deformation and shape effects in hot nucleus, where our present formalism works very well as seen in our earlier works ~\cite{MAPLB10,MANPA19-23} which showed good agreement with the experimental data proving the efficacy and reliability of this methodology.\par

\begin{figure}[h]
\centering
\begin{subfigure}[t]{0.255\textwidth}
  \centering
  \includegraphics[width=\linewidth, height=10cm]{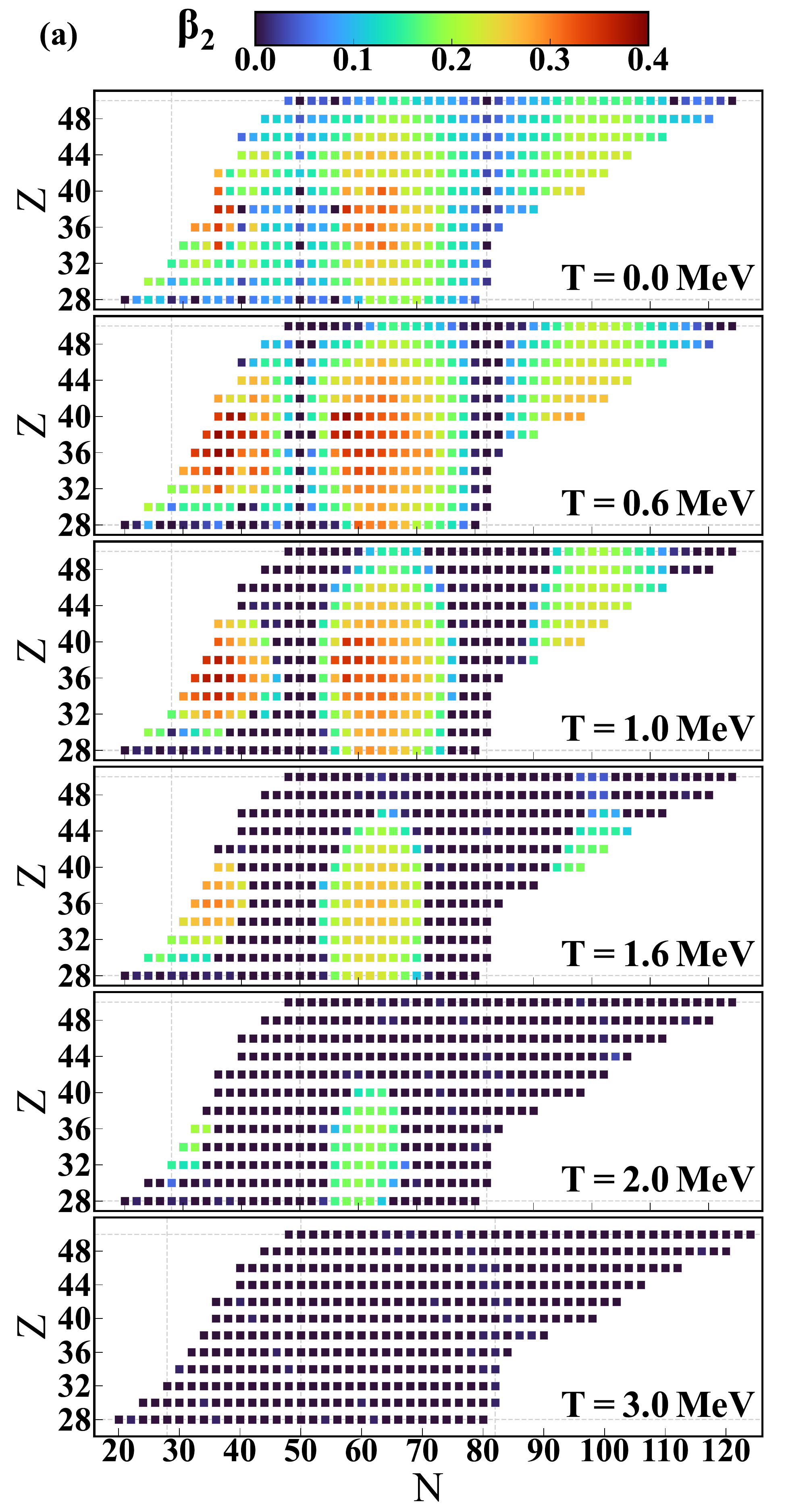}
  \label{fig:beta2_def}
\end{subfigure}\hspace{-4 pt}%
\begin{subfigure}[t]{0.22\textwidth}
  \centering
  \includegraphics[width=\linewidth, height=10cm]{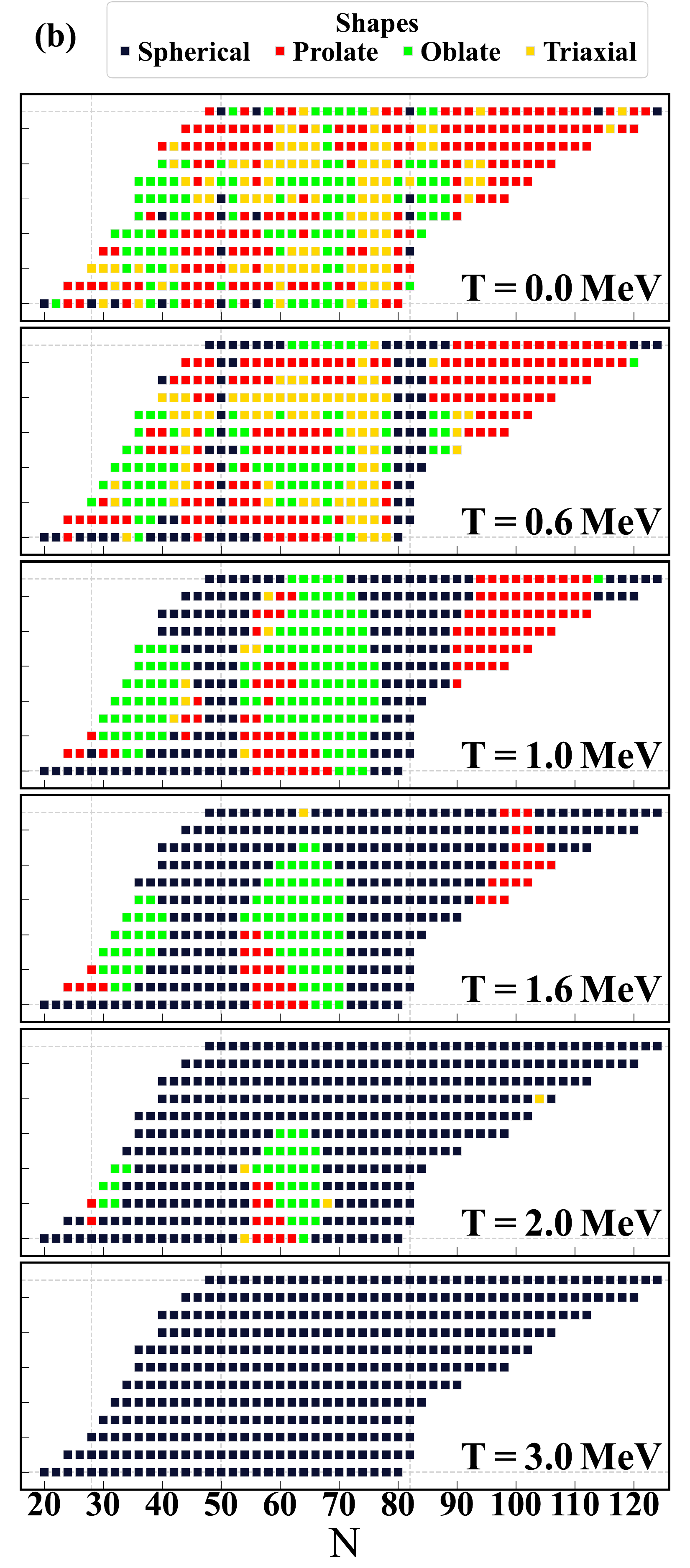}
  \label{fig:gamma_shape}
\end{subfigure}
\captionsetup{skip=-5pt}
\caption{Temperature dependence of nuclear deformation parameters for $Z = 28$--$50$: (a) Variation of the axial deformation parameter ($\beta_2$) and (b) Evolution of the nuclear shape characterized by the triaxial deformation parameter ($\gamma$).}

\label{fig:BetaGamma}
\end{figure}

Figure~\ref{fig:BetaGamma} presents the thermal evolution of nuclear deformation (Fig. 2(a)) and shapes (Fig. 2(b)) as predicted by Nilsson deformation parameters ($\beta_2$, $\gamma$), displayed on the $N$--$Z$ plane for isotopic chains of even-even nuclei Z $=$ 28 to 50 for ground state (T = 0 MeV) and T $=$ 0.6 $-$ 3.0 MeV.  At zero temperature, extensive regions of the chart exhibit finite deformation, with prolate, oblate, and triaxial shapes  coexisting across the wide $N-Z$ landscape except at magic nuclei with shell closures at N = 50 and 82 indicating pronounced shell effects where the spherical shape dominates. As temperature rises to $T \approx 0.6$ -- $1.0$~MeV, the shell effects weaken, $\beta_2$ reduces and the nuclear shapes evolve (as seen in Fig. 2(b)) where many nuclei transition from triaxial to axially symmetric, while some transition directly to spherical from prolate, oblate and triaxial shapes. It is interesting to see the triaxial shapes being almost washed out at T = 1 MeV with transition to either spherical or axially symmetric shapes indicating that the triaxial phase is more common in low temperature states. At higher T, the thermal excitations gradually overcome the mean-field forces that sustain deformation, resulting in a sharp reduction in deformation ($\beta_2$) at  T $\approx$ 1 $–$ 2 MeV. At a characteristic critical temperature ($T_c$), the $\beta_2$ values tend to zero and the nucleus transitions to a nearly spherical shape, indicating the shell quenching effect visible in Fig. \ref{fig:BetaGamma} where the $(\beta_2,\gamma)$ plane show zero deformation (in left panel) and spherical shapes (in right panel) at $T=2-3$ MeV except for very few deformed states seen at T = 2 MeV with oblate and prolate configurations in mid shell region. The vanishing of nuclear deformation with increasing temperature can be understood microscopically as a consequence of the thermal smearing of the Nilsson single-particle occupations (Eqs. (5$-$7) in Supplementary Note 2), which weaken the deformation-induced shell gaps near the Fermi surface and leads to a restoration of spherical level degeneracy at the critical temperature, as shown in our previous works (Fig. (3) of Ref. \cite{MAPLB10}, and (Figs. 7 and 8) of Ref. \cite{MAPRC04}).
\begin{figure}[h!]
	\centering 
	\includegraphics[width = 0.495\textwidth]{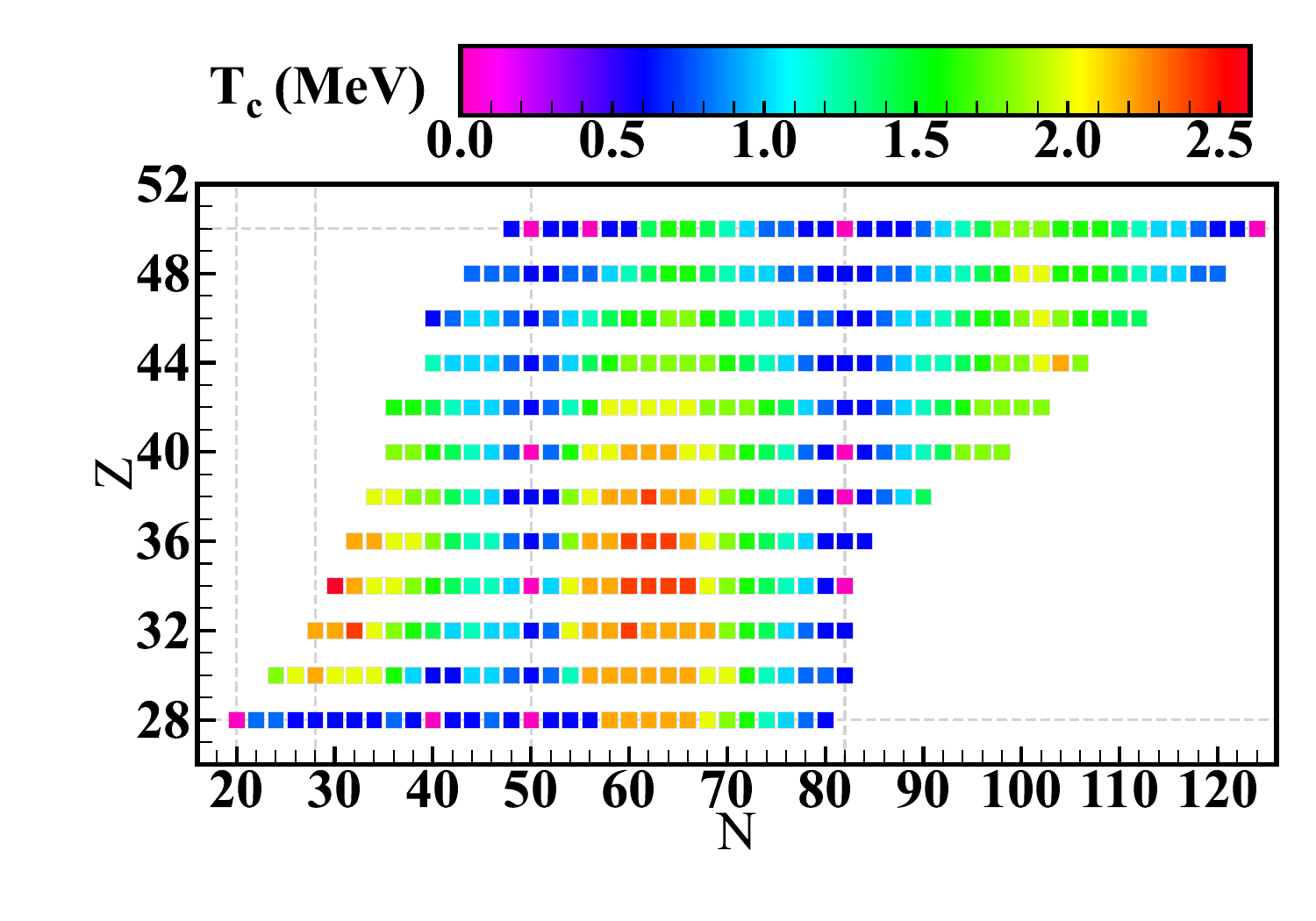}	
	\captionsetup{skip=-5pt}
	\caption{Distribution of the critical temperature $T_c$ in the $N$–$Z$ plane for even–even nuclei with $Z = 28$–$50$. The color scale indicates the temperature at which deformation highlighting the role of shell structure, with reduced $T_c$ near the neutron shell closures at $N = 50$ and $82$.} 
	\label{fig3-Tc}
\end{figure}

Fig. \ref{fig3-Tc} displays the distribution of critical temperature $T_c$ for all considered even–even nuclei in the $N$–$Z$ plane. The  observed color gradients in the $N$–$Z$ chart reveal a pronounced dependence of $T_c$ on shell structure where the thermal response is not uniform across the nuclear landscape but is strongly modulated by underlying single-particle structure. Nuclei located near major shell closures at 50 and 82 exhibit systematically lower critical temperatures ($<$ 1 MeV), reflecting the lesser deformed or nearly spherical states. In contrast, mid-shell and nearby nuclei show higher $T_c$ values around 1.5$-$2.0 MeV in most of the nuclei that are well deformed. The critical temperature $T_c$ is different for different nuclei and is typically nucleus-dependent ~\cite{GoodmanNPA81}. These findings have important implications for nuclear level density ~\cite{MANPA19-23} models and the statistical description of compound nuclear reactions at high excitation energies.\par

\begin{figure}[h!]
	\centering 
	\includegraphics[width = 0.49\textwidth]{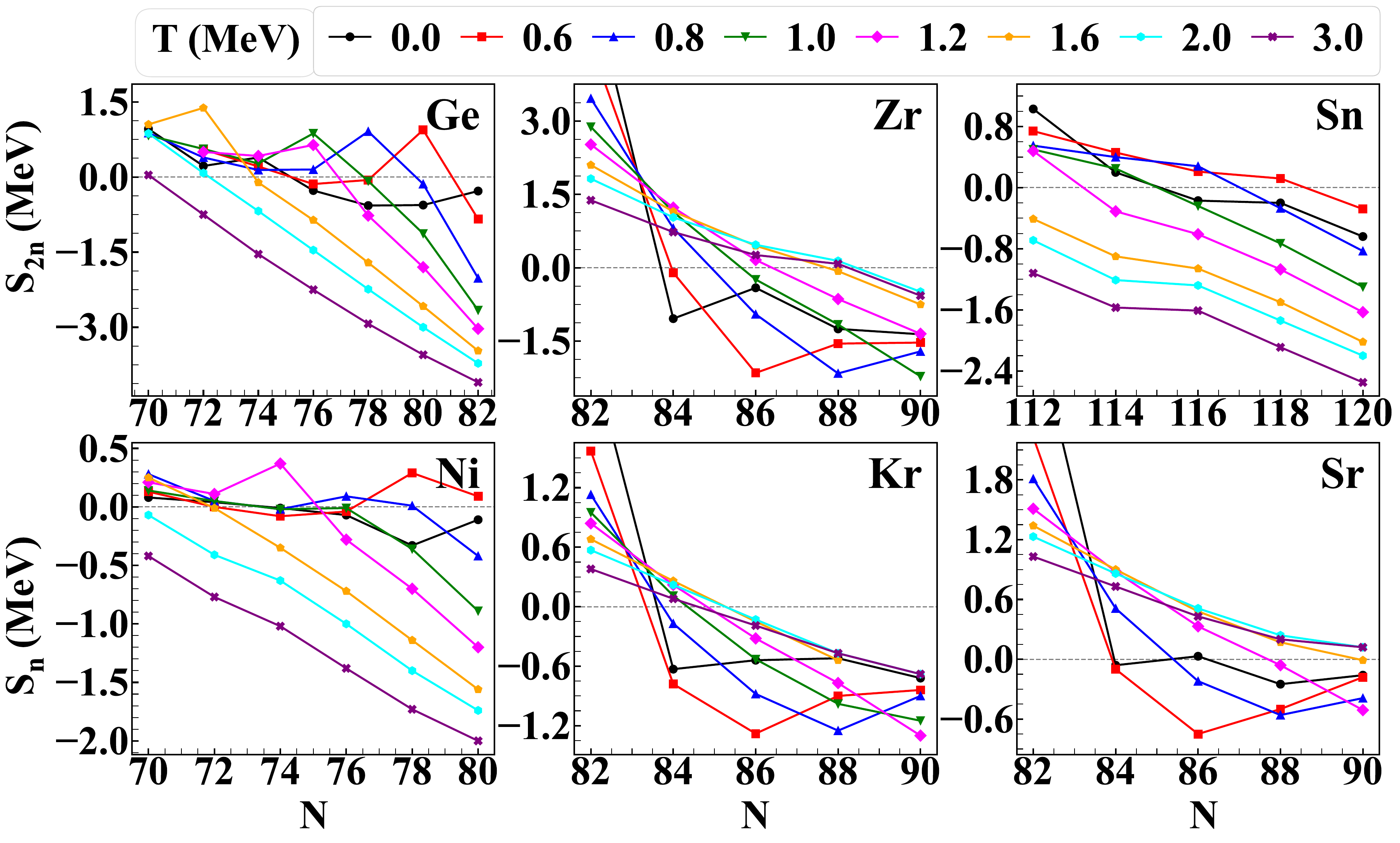}	
	\caption{Variation of two neutron separation energy ($S_{2n}$) and neutron separation energy ($S_n$) with temperature for nuclei exhibiting dripline expansion.} 
	\label{SnS2n}%
\end{figure}

\begin{figure}[h!]
	\centering 
	\includegraphics[width = 0.495\textwidth]{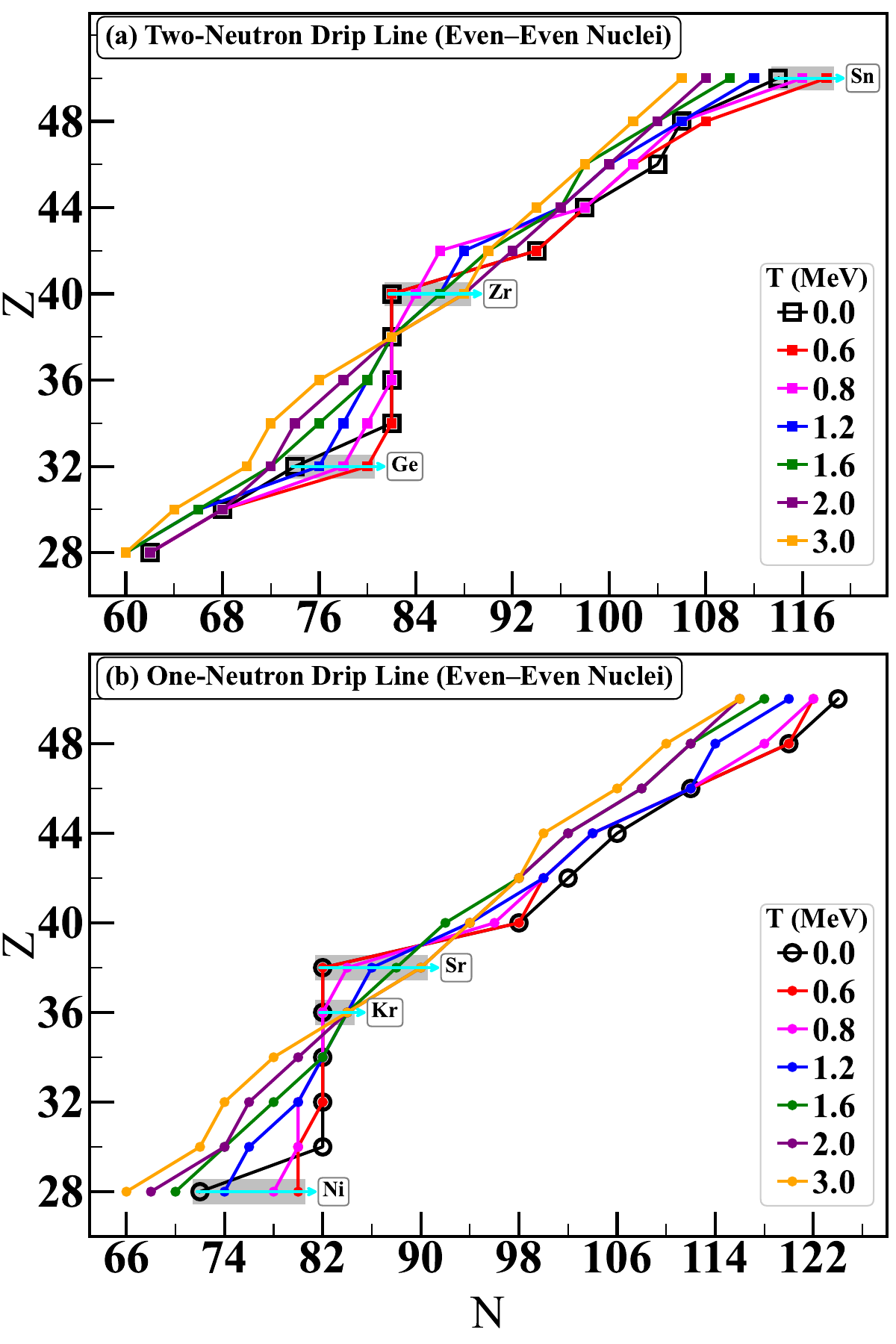}	
	\caption{Evolution of the (a) two-neutron and (b) one-neutron drip lines for even–even nuclei with $Z = 28$–$50$ as a function of temperature ($T = 0$ and $0.6$–$3.0$ MeV). With increasing temperature, selected isotopic chains exhibit a systematic shift of the drip lines toward higher neutron numbers marked with arrow. This behavior correlates with the temperature-induced quenching of nuclear deformation and the transition toward near-spherical shapes, which modifies the binding of weakly bound neutrons.} 
	\label{Fig_2n_1n_dripline.eps}%
\end{figure}

As thermal effects reduce shell gaps and enhance the level density and entropy, the weakening of binding is pronounced at finite temperatures, affecting the nuclear stability which is best reflected in the nucleon separation energies. We trace nucleon drip lines using our computed proton and neutron separation energy as a function of temperature T (=  0.6 - 3.0 MeV). Usually the separation energy decreases with increasing temperature due to thermal excitations which populate higher single particle levels, the shell effects get weakened that reduces the gain of binding from shells implying the reduced binding in hot nuclei. But at certain instances, we observed that with increasing temperature, the separation energy shows some unusual enhancement by few keVs which is found to be closely correlated with a structural change with a sharp decline or collapse in $\beta_2$ with increasing temperature. When the deformation collapses or reduces significantly with increasing temperature, the occupation probabilities smear and the single-particle levels reorganize, the deformation induced levels near the Fermi surface may redistribute into the spherical level degeneracy, effectively lowering their energy relative to the Fermi surface. This redistribution of levels may result in locally enhancing the binding that makes the removal of a nucleon slightly harder (enhancing the separation energy) that sometimes shift the last unbound nucleon to the bound one and contribute to a modest extension of the proton and neutron drip lines which has been observed in neutron rich nuclei 
$^{100-108}\mathrm{Ni},\ ^{106-112}\mathrm{Ge},\ ^{118-120}\mathrm{Kr},\ ^{120-128}\mathrm{Sr},\ ^{124-128}\mathrm{Zr},\ ^{164-168}\mathrm{Sn}$
 demonstrated in Figs.~\ref{SnS2n} and \ref{Fig_2n_1n_dripline.eps}. \par

Fig.~\ref{SnS2n} displays the $S_{2n}$ and $S_n$ for neutron rich isotopes of  Ni, Ge, Kr, Sr, Zr and Sn lying close to drip line. It is interesting to note that some isotopes, specifically  $^{102-108}\mathrm{Ni},\ ^{108-112}\mathrm{Ge},\ ^{118}\mathrm{Kr},\ ^{122-128}\mathrm{Sr},\ ^{124-128}\mathrm{Zr},\ ^{166-168}\mathrm{Sn}$
 with negative separation energy in ground state or at low T, become weakly bound with positive separation energy with increasing temperature, thereby expanding the drip line. 

Fig.~\ref{Fig_2n_1n_dripline.eps} illustrates the thermal evolution of two-neutron drip line ($S_{2n} \le 0$) (Fig. 5(a)) and one-neutron drip line ($S_{n} \le 0$) (Fig. 5(b)) in N-Z plane for even–even isotopes of $Z = 28$--$50$ for ground state (T = 0 MeV) and temperatures $T = 0.6$ to $3.0$~MeV. Drip lines exhibit a pronounced dependence on shell structure and deformation at low T with clear signatures near the neutron shell closure at $N = 82$. A notable example is provided by the Zr isotopic chain, where the two-neutron drip line extends from $N = 82$ at $T = 0$ to $N \approx 88$ at $T = 3.0$~MeV. Similar, though less pronounced, extensions are observed at intermediate temperatures for the Ge and Sn isotopes. These trends are qualitatively consistent with results obtained within the relativistic Hartree–Bogoliubov (RHB) framework and its finite-temperature extension (FT-RHB), employing different covariant energy density functionals such as DD-PC1 and DD-PCX, and including the proper treatment of continuum effects through the BLV subtraction procedure \cite{RavlicNC23}.

The lower panel (Fig. 5(b)) shows that the one-neutron drip line exhibits analogous temperature dependence for certain isotopic chains, most prominently for Ni, Kr and Sr. For these nuclei, systems that are unbound at zero temperature due to finite deformation and negative neutron separation energies become marginally bound at higher temperatures.\par

As the temperature increases, usually the neutron drip line shifts to lower neutron numbers except for few isotopes (as seen in Fig.5 (a) and (b)) which show an unusual shift of the drip lines toward larger neutron numbers (marked with an arrow sign along with the nucleus name mentioned on its side). This behavior is consistent with the temperature-driven reduction of deformation discussed in Fig.~\ref{fig:BetaGamma}, which leads to near-spherical configurations and modified binding properties that results in an unbound nucleus to become a bound state. This analysis shows an uncommonly seen non-trivial correlation between the thermal shape fluctuations and separation energy that influences the precise position of the drip line which is an important outcome of the present work. \par

These findings generate further curiosity and impetus for a deeper study on the thermal response to the dynamics of other decay modes when the nucleon emission is suppressed and we find many nuclei studied here that de-excite by $\beta$-decay which may be relevant in stellar processes.  Although $\beta$-decay is not the dominant weak-interaction process in all astrophysical environments, but during the pre-collapse stellar burning, hot freeze-out conditions, and the high-temperature phases preceding the onset of \textit{r}-process nucleosynthesis, temperatures of the order $T \gtrsim 1~\mathrm{MeV}$ can be achieved \cite{mumpower2016,arcones2013,qian2007}, where the temperature-induced changes in nuclear deformation and binding energies may influence neutron separation energies and  weak-interaction rates, thereby indirectly affecting subsequent nucleosynthesis pathways \cite{rauscher2013}.\par
  
To investigate the implications of structural changes induced by temperature on weak interaction rates, we study the influence of nuclear deformation on Q-values associated with $\beta$-decay in hot nuclei. Theoretical descriptions of $\beta$-decay lifetimes employ a variety of models, each designed to capture specific aspects of nuclear dynamics. Since $\beta$-decay is governed by the weak interaction, the lifetime is determined primarily by the available lepton phase space and the strength of the nuclear transition matrix elements. These transitions are classified as allowed or forbidden depending on the angular momentum and parity change, and their reliable treatment often requires shell-model or QRPA-based approaches to incorporate configuration mixing and pairing correlations \cite{Wilson68,BehrensNPA162,CaurierRMP77,SarrigurenPRC93}. While such microscopic methods yield valuable insight, their large-scale application, especially in systematic studies involving finite temperature, is computationally demanding. For this purpose, Q-value based semi-empirical formulas provide a simpler yet effective alternative. Derived from global fits to experimental data, these expressions capture essential structural effects implicitly and allow practical half-life estimates across the nuclear chart. In view of the above, we apply a semi-empirical formula \cite{SobhaniCJP23} to compute both $\beta^+$ and $\beta^-$ - decay lifetimes at finite temperatures in the present work. The approach shows good accuracy, with RMSE values of about 1.00 sec. for a dataset of 945 $\beta^+$ emitters and 0.74 sec. for 994 $\beta^-$ emitters from NUBASE2020 \cite{WangCPC45}, when benchmarked against established microscopic frameworks such as FRDM+QRPA \cite{MollerANDT12} and RHB+RQRPA \cite{MarketinPRC93}. To assess the reliability of the temperature dependence of the calculated $\beta$-decay lifetimes, we compare our results with available finite-temperature QRPA calculations as can be seen in the Supplementary Note 2. The consistent trends observed across different microscopic approaches indicate that the present method captures the essential thermal effects on weak decay rates. Although there remains considerable scope for further refinement, the results demonstrate that such formulas offer a concrete means to assess lifetime variations driven by Q-value changes due to structural transitions at elevated temperatures. \par

\begin{figure}[h!]
	\centering 
	\includegraphics[ width=0.5\textwidth]{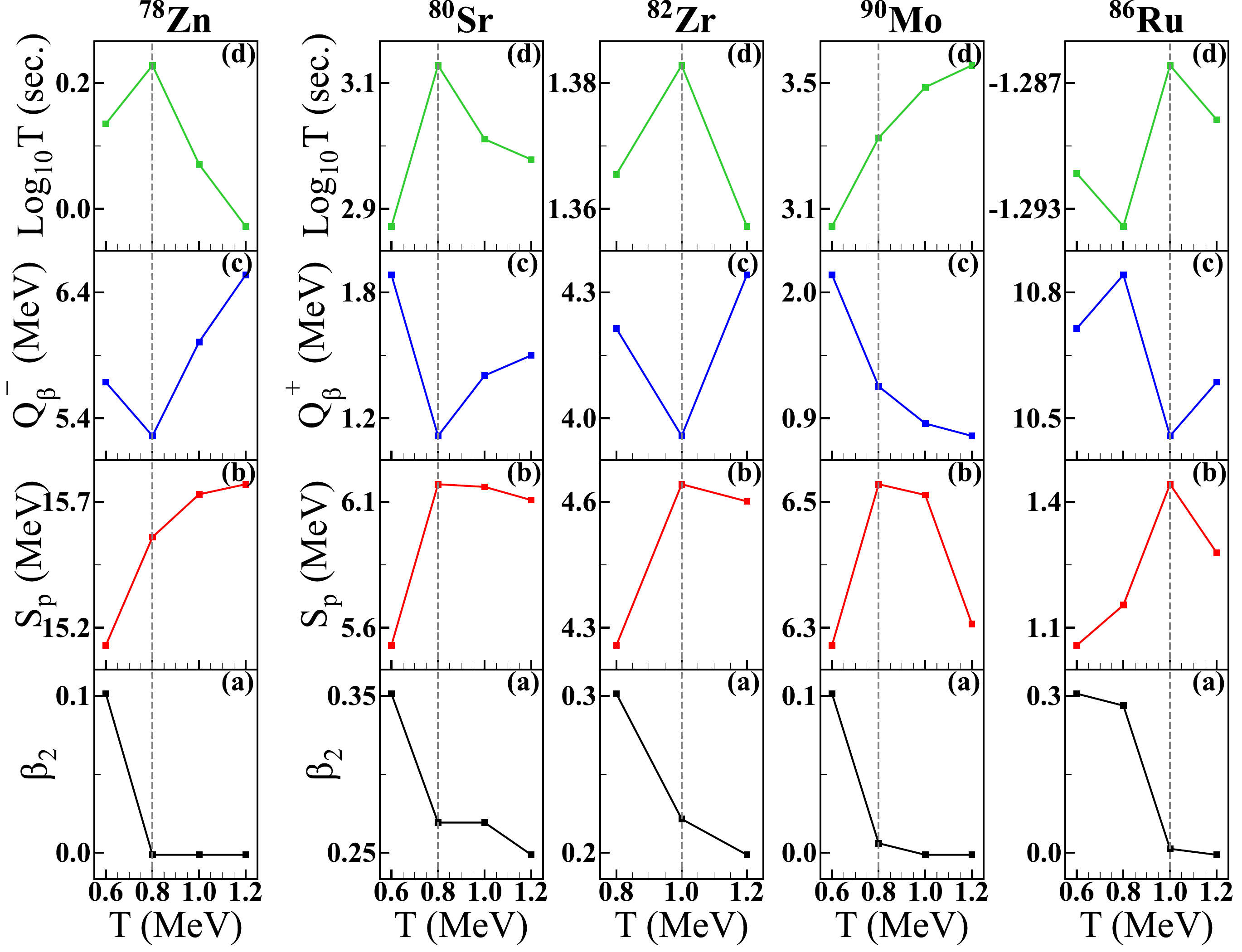}
	\caption{Temperature-induced evolution in (a) deformation ($\beta_2$), and its impact on  (b) proton separation energy ($S_p$), (c) Q-value ($Q_{\beta^-}$) for $^{78}$Zn exhibiting $\beta^-$- decay, and  Q-value ($Q_{\beta^+}$) for $^{80}$Sr, $^{82}$Zr, $^{90}$Mo, and $^{86}$Ru, exhibiting  $\beta^+$- decay, and their corresponding (d)  $\beta$-decay half-lives ($\log_{10}T$).}
	\label{fig:SpLifetimeZ30-44}
\end{figure}
\begin{figure}[h!]
	\centering 
	\includegraphics[width=0.5\textwidth]{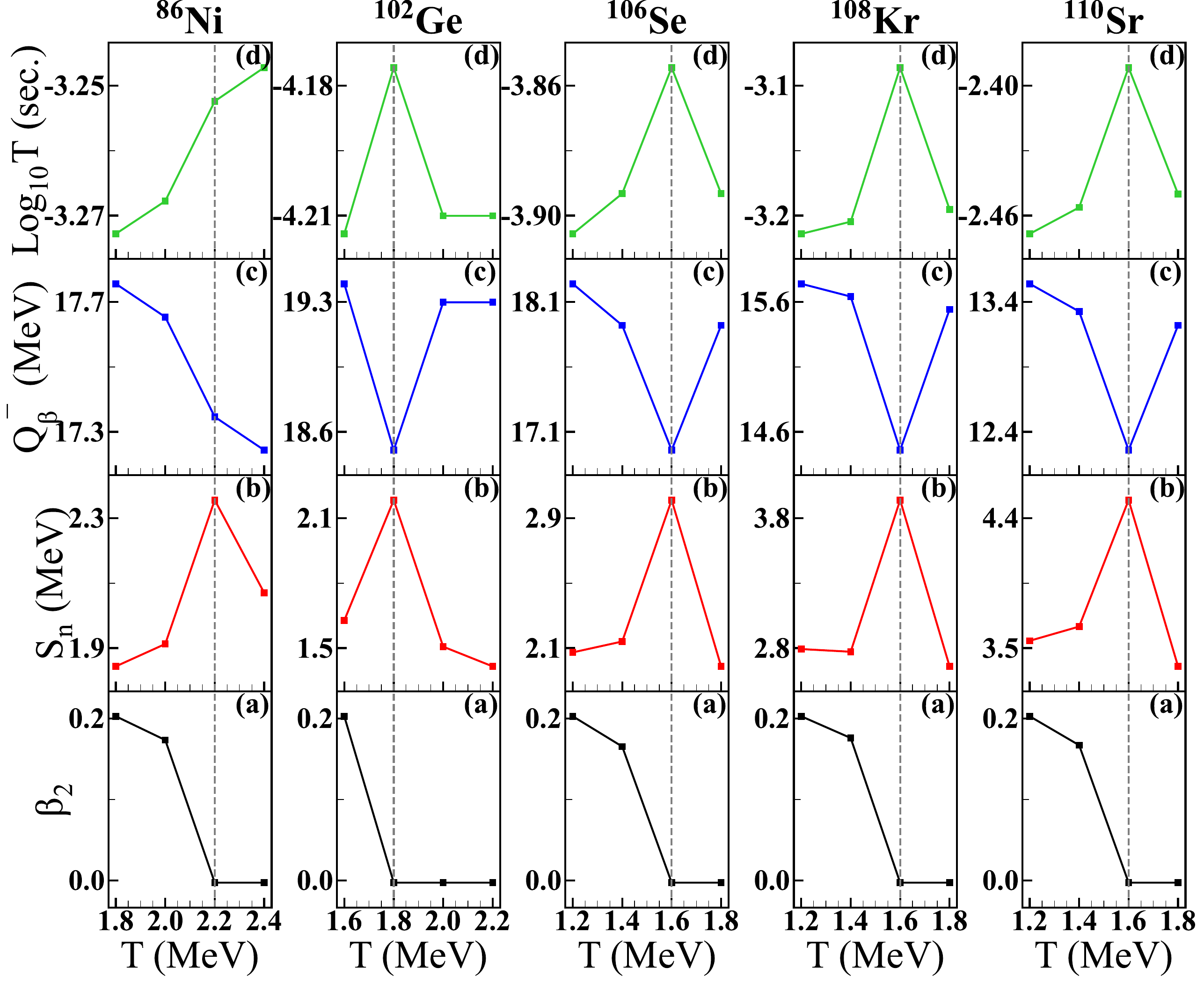}
	\caption{Temperature-induced evolution in (a) deformation ($\beta_2$), which impacts the (b) neutron separation energy ($S_n$), (c) $\beta^-$- decay $Q$-value ($Q_{\beta^-}$), and (d) $\beta$-decay half-lives ($Log_{10}T$).}
	\label{fig:SnLifetimeZ28-38}
\end{figure}
\begin{figure}[h!]
	\centering 
	\includegraphics[ width=0.5\textwidth]{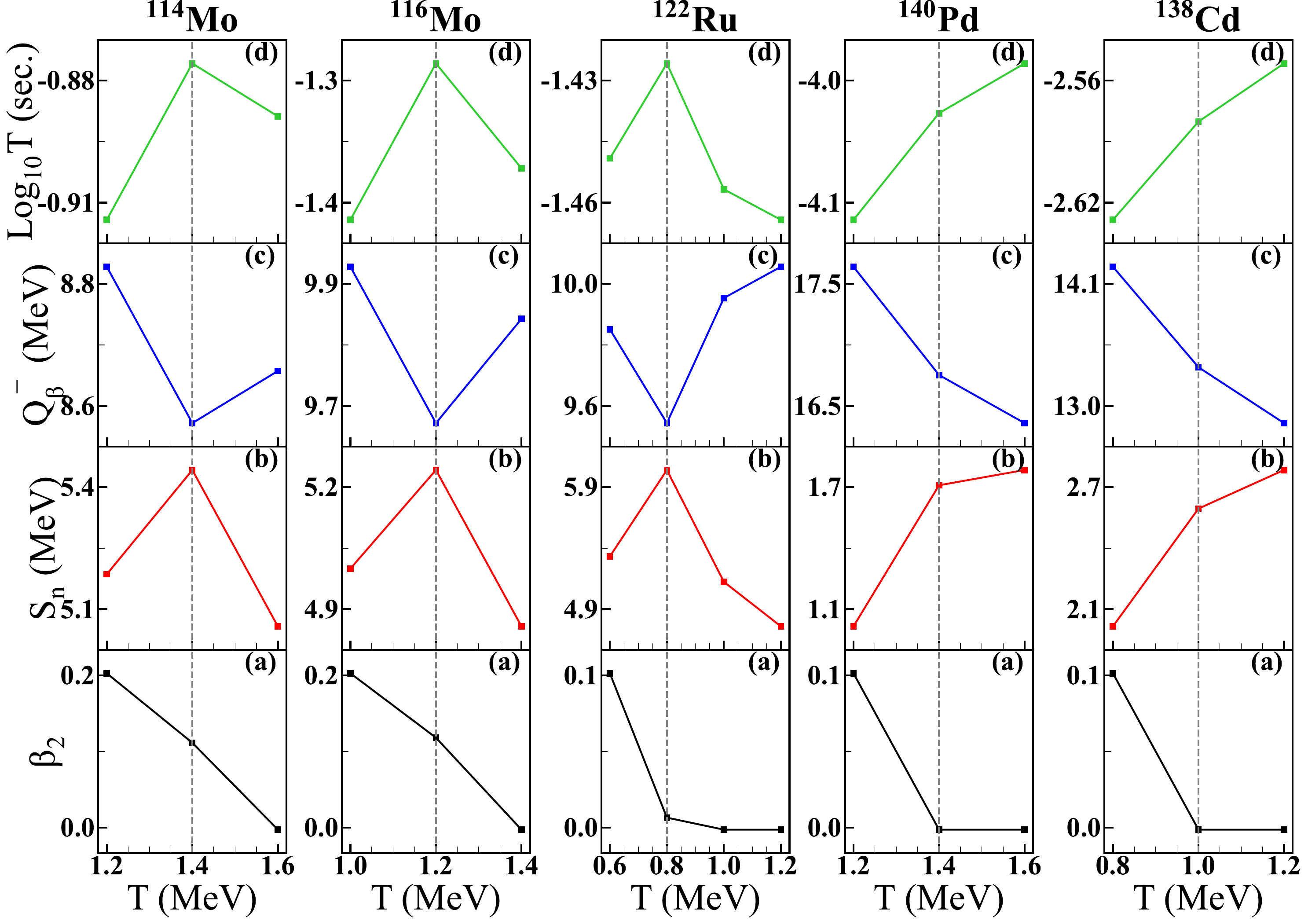}
	\caption{Temperature-induced evolution in (a) deformation ($\beta_2$), which impacts the (b) neutron separation energy ($S_n$), (c) $\beta^-$- decay $Q$-value ($Q_{\beta^-}$), and (d) $\beta$-decay half-lives ($Log_{10}T$).}
	\label{fig:SnLifetimeZ42-48}
\end{figure}

Around the critical temperatures where the deformation decreases and $S_{p}$ or $S_n$ increases (Figs. \ref{fig:SpLifetimeZ30-44} -- \ref{fig:SnLifetimeZ42-48}), we observe a corresponding decrease in $Q_\beta$ value that results in the reduced decay rates. The observed correlation between decreasing deformation and increasing $S_p$ or $S_n$, along with the variations in $Q_\beta$ and lifetime, highlights a complex and previously unexplored interplay between shape evolution, binding energy, and $\beta$-decay behaviour at finite temperature.\par

Fig.~\ref{fig:SpLifetimeZ30-44} shows variation in deformation parameter ($\beta_2$), proton separation energy ($S_p$), ~$\beta$-decay Q-value ($Q_{\beta^\pm}$), and ~$\beta$-decay lifetime ($\log_{10} T$) as a function of temperature for nuclei $^{78}$Zn $^{80}$Sr, $^{82}$Zr, $^{90}$Mo, and $^{86}$Ru. $\beta$-decay Q-value plot shows $Q_{\beta^-}$ - decay mode for $^{78}$Zn exhibiting $\beta^-$ - decay, and $Q_{\beta^+}$ - decay mode for neutron deficient $^{80}$Sr, $^{82}$Zr, $^{90}$Mo, and $^{86}$Ru, exhibiting $\beta^+$ - decay. In this figure (Fig.~\ref{fig:SpLifetimeZ30-44}) we observed some very interesting interplay of structural effects and decay processes not shown before in any other work as far to our knowledge. \par

%
In Fig.~\ref{fig:SpLifetimeZ30-44}, panel (a) for each nucleus shows a significant drop in deformation with increasing $T$ between 0.6 to 1.2 MeV which is accompanied by an increase in the proton separation energy ($S_p$) by a few keVs providing more stability to the nucleus. At this transition, the $Q_\beta$ value  also shows a slight  decline  leading to a modest increase in the $\beta$-decay lifetime. This shows a very significant impact of shell-quenching effects on particle stability by the enhancement in proton separation energy
($S_p$), decline in $Q_\beta$ resulting in higher $\beta$-decay lifetime. This
enhancement in lifetime may slow down $\beta$-decay process and
may provide enough lifetime for probes. \par

In  some cases, as observed for $^{80}$Sr and $^{82}$Zr, a small decline in deformation also causes an increase in $S_p$, a lower $Q_{\beta^+}$ value accompanied by an enhancement in ${\beta^+}$ - decay lifetime which emphasizes that sometimes even a  small deformation fluctuation can influence the binding and decay lifetimes. It is important to note here that these nuclei are not lying at the drip line and hence they do not contribute to drip line expansion, but in principle, the structural change brings more stability to the nucleus by enhancing the separation energy, which in case of drip line nuclei causes drip line expansion. \par 


Figs.~\ref{fig:SnLifetimeZ28-38} and \ref{fig:SnLifetimeZ42-48} exhibit deformation parameter ($\beta_2$), along with neutron separation energy ($S_n$), $Q_{\beta^-}$ value, and $\beta$-decay life-time ($\log_{10}T$) for stable and neutron rich nuclei $^{86}$Ni, $^{102}$Ge, $^{106}$Se, $^{108}$Kr, $^{110}$Sr, $^{114}$Mo, $^{116}$Mo, $^{122}$Ru, $^{140}$Pd, and $^{138}$Cd around the critical temperatures. A similar interplay of structural and decay
properties is found in these isotopes around $T_c$ ranging between 1.2 --
2.2 MeV. A sharp decline in deformation from a high $\beta_2$ value $\geq$ 0.2 to zero value is observed. At this transition,  $S_n$ rises and $Q_{\beta^-}$ decreases that leads to a small enhancement of the $\beta^-$ - decay lifetime.
With further increasing T, we notice that the deformation remains at zero and the separation energy normally reduces with increasing T, $Q_{\beta^-}$ increases and its lifetime reduces sharply, which is the expected usual behaviour with the rise in temperature. The unexpected enhancement in neutron separation energy observed here, and the decline in $Q_{\beta^-}$ value occur with the sudden and significant drop in deformation due to shell quenching effects in most cases. The enhancement of separation energy provides extra stability, and the $\beta$-decay may be delayed due to enhancement of lifetime which shows significant impact of temperature induced structural changes on decay properties, which is a major highlight of this work.  \par


The correlations between the shell structure and lifetimes in hot nuclei imply that the $\beta$-decay lifetimes are sensitive to the structural transitions at all excitations, but the unusual enhancement in binding, a decrease in $Q_\beta$, and the deceleration of $\beta$-decay occurs only at certain temperatures where significant structural changes appear either at low temperatures where the  shell effects are prominent or  around the critical temperatures $T_c$ where the shell quenching effects are  predominant. The trend of variation is very similar in all the nuclei studied here highlighting the impact of temperature induced shape changes in decay processes in hot nuclei, which may influence the thermonuclear reactions in stellar environments. These findings emphasize the need for nuclear structure models to include finite-temperature shape effects in view of the importance of nuclear structure in astrophysical nucleosynthesis modeling.\par

\section{Conclusion}

To conclude, we have investigated the role of microscopic structural effects and finite-temperature dynamics on particle stability and nuclear decay properties of hot nuclei across the isotopic chains of $Z = 28$–$50$, and discussed their potential implications for astrophysical environments. Using the  statistical theory of hot nuclei combined with macroscopic-microscopic approach including Nilsson $-$ Strutinsky Method, we systematically examined the thermal response of nuclear shapes, shell effects, and decay-related observables in a broad region of the nuclear chart that is relevant for nucleosynthesis pathways and extreme stellar conditions.

Our results show that increasing temperature leads to a gradual quenching of deformation, driving nuclei toward near-spherical configurations at characteristic critical temperatures $T_c \sim 1$–$2$ MeV. This behavior reflects the progressive weakening of shell effects under thermal excitation and exhibits a clear dependence on neutron and proton shell structure. The calculated neutron and proton separation energies generally decrease with temperature, indicating reduced binding in hot nuclear systems. However, in selected isotopes, a pronounced reduction in deformation  is accompanied by a temporary enhancement of separation energies, resulting in increased stability and, in some cases, a shift of the one- or two-neutron drip lines toward more bound configurations.

A remarkably strong correlation between thermal shape evolution and the temperature dependence of calculated $Q_\beta$ values and $\beta$-decay lifetimes has been shown. These results indicate that microscopic structural changes in hot nuclei can directly influence weak decay properties, thereby modifying decay timescales under finite-temperature conditions. While the present study does not aim to provide direct astrophysical rate calculations, it highlights the importance of incorporating temperature-dependent nuclear structural effects into models of nucleosynthesis and stellar evolution.

Overall, this work underscores the need for coordinated theoretical developments, including finite-temperature microscopic approaches, together with future experimental constraints and astrophysical modeling, to achieve a consistent description of nuclear structure and decay in hot and dense stellar environments.

\section{Acknowledgement}
M.A. and G.S. acknowledge support from DST, Govt. of India, and DST, Govt. of Rajasthan, Jaipur, India under WIDUSHI-B/PM/2024/23 and F24(1)DST/R$\&$D/2024-EAC-00378-6549873/819, respectively.\par

\bibliographystyle{elsarticle-num}

\end{document}


\begin{center}
{\Large \textbf{Supplementary Information}}\\[6pt]

\textbf{Title:} Impact of microscopic structural transitions on particle stability and lifetimes of hot nuclei\\
Mamta Aggarwal, Pranali Parab and G. Saxena
\end{center}
\textbf{}		
\subsection*{Supplementary Note 1.: Theoretical Formalism}
The excited states of the nucleus are treated using the statistical theory~\cite{MTN,MNV} with the grand canonical partition function 
Q($\alpha$,$\beta'$) of the deformed nuclear system of N neutrons and Z protons  
\begin{equation}
Q(\alpha_Z,\alpha_N,\beta')=\sum \exp(-\beta' E_i+ \alpha_Z Z_i+\alpha_NN_i)
\end{equation}
The Lagrangian multipliers $\alpha$ and $\beta'$ project the 
particle number and total energy of the system in an 
approximate manner through the saddle point equations at temperature T(=1/$\beta'$) 
\begin{equation}
- \partial \ln Q/\partial \beta' = <E>
\end{equation}
\begin{equation}
 \partial \ln Q / \partial \alpha_Z = <Z>
\end{equation}
\begin{equation}
\partial \ln Q/\partial \alpha_N = <N>
\end{equation}

The corresponding equations in terms of single-particle eigen values for the protons $\epsilon_i^Z$ and neutrons $\epsilon_i^N$ ~\cite{HM} of the deformed oscillator potential of the Nilsson Hamiltonian diagonalized with cylindrical basis states ~\cite{EIS, GS} are given as
\begin{equation}
 < Z >  = \sum n_i^Z = \sum [1 + \exp(-\alpha_Z +\beta' \epsilon_i^Z)]^{-1},
\end{equation}
\begin{equation}
 < N >  = \sum n_i^N = \sum [1 + \exp(-\alpha_N +\beta' \epsilon_i^N)]^{-1},
\end{equation}
\begin{equation}
 <E(T)>  = \sum n_i^Z \epsilon_i^Z + \sum n_i^N \epsilon_i^N .
\end{equation}
where, $n_i^Z$ and $n_i^N$ are occupation probabilities for protons and neutrons respectively.

\footnotetext[1]{The Lagrangian multiplier $\beta'$ should not be confused with the deformation parameter $\beta_2$. The prime is used only to differentiate it from the deformation parameter. The notation $\beta'$ used in the text from Eqs. (1), (2), (5) and (6) represents the Lagrangian multiplier, whereas elsewhere in this article $\beta_2$ denotes the deformation parameter.}

The excitation energy of the system is found by 
\begin{equation} 
 E^*(T) = E(T) - E(0,0)
\end{equation}
where E(0,0) is the ground state energy of the nucleus given by 
\begin{equation}
E(0,0) = \sum \epsilon_i^Z + \sum \epsilon_i^N .
\end{equation}

The entropy of the system is found by 
\begin{equation}
S  = -\sum [n_i \ln n_i + (1- n_i) \ln(1- n_i)] .
\end{equation}

To evaluate the separation energies of an excited nucleus as a function of temperature, we first calculate the binding energy and the ground state separation energy 
using the macroscopic-microscopic approach as done earlier~\cite{MAM}.
\begin{eqnarray}
BE_{gs}(Z,N,\beta_2,\gamma) = BE_{LDM}(Z,N) -  \nonumber\\
 E_{def}(Z,N,\beta_2,\gamma)-\delta E_{shell}(Z,N,\beta_2,\gamma)
\end{eqnarray}
where the macroscopic binding energy BE$_{LDM}$ is obtained from the LDM mass formula of Moller-Nix~\cite{PM} which reproduces the binding energies quite well over a wide range of nuclei. The microscopic effects arising due to nonuniform distribution of nucleons are included through the Strutinsky's shell correction $\delta$E$_{shell}$ ~\cite{VM} along with the deformation energy E$_{def}$ (obtained from the surface and Coulomb effects). Shell correction to energy $\delta$E$_{shell}$ can be written as 
\begin{equation}
\delta E_{shell}=\sum_{i=1} ^A \epsilon_i- \tilde E
\end{equation}
the first term being the shell model energy in the ground state and the second term is the smoothed energy with the smearing width 1.2$\hbar \omega$. The smearing is done by the Gaussian distribution function (see Ref.~\cite{MAMI} for details). The difference between the binding energies 
BE$_{gs}$ of the parent and daughter nuclei gives the corrected ground state separation energy.\\

Total energy E (= -BE) of a nucleus at finite temperature T is calculated by blending the macroscopic-microscopic approach and statistical theory by  
summing the ground state energy obtained from the macroscopic-microscopic approch (Eq.(11)) with the total excitation energy E$^*$ (Eq.(8)) obtained from the statistical theory as  
\begin{eqnarray}
E(Z,N,\beta_2,\gamma,T)= - BE(Z,N,T,\beta_2,\gamma)\nonumber\\
  = - BE_{gs}(Z,N) + E^*(T,\beta_2,\gamma)\nonumber\\
  = - BE_{LDM}(Z,N) + E_{def}(Z,N,\beta_2,\gamma)\nonumber\\
 +\delta E_{shell}(Z,N,\beta_2,\gamma)+ E^*(Z,N,\beta_2,\gamma,T)
\end{eqnarray}
Free energy F is minimized~\cite{MAI} with respect to the Nilsson deformation 
parameters $\beta_2$ and $\gamma$ which gives the shape and deformation of the 
excited nucleus at a fixed temperature T.
\begin{eqnarray}
F(Z,N,\beta_2,\gamma,T) = E(Z,N,\beta_2,\gamma,T) \nonumber\\ 
- T * S(Z,N,\beta_2,\gamma,T)
\end{eqnarray}
The above calculations are performed for the angular deformation parameter $\gamma$ in the range 180$^o$ (oblate) to -120$^o$ (prolate) with triaxial shapes lying in between and the axial deformation parameter $\beta_2$ range from 0 to 0.4 in the steps of 0.01.\\

One proton separation energy of the excited nucleus is obtained as 
\begin{eqnarray}
S_P(Z,N,\beta_2,\gamma,T)= BE(Z,N,\beta_2,\gamma,E^*,T) - \nonumber\\
 BE(Z-1,N,\beta_{2_d},\gamma_d,E^*_d,T )
\end{eqnarray}

Two proton separation energy of the excited nucleus is obtained as 
\begin{eqnarray}
S_{2P}(Z,N,\beta_2,\gamma,T)= BE(Z,N,\beta_2,\gamma,E^*,T) - \nonumber\\
 BE(Z-2,N,\beta_{2_d},\gamma_d,E^*_d,T )
\end{eqnarray}
where E$^*$ and E$^*_d$ are the excitation energies of the parent and corresponding daughter nuclei respectively.\\
Similarly, the one neutron and two neutron separation energies of the excited nucleus are obtained as
\begin{eqnarray}
S_n(Z,N,\beta_2,\gamma,T)= BE(Z,N,\beta_2,\gamma,E^*,T) - \nonumber\\
BE(Z,N-1,\beta_{2_d},\gamma_d,E^*_d,T )
\end{eqnarray}
and
\begin{eqnarray}
S_{2n}(Z,N,\beta_2,\gamma,T)= BE(Z,N,\beta_2,\gamma,E^*,T) - \nonumber\\
BE(Z,N-2,\beta_{2_d},\gamma_d,E^*_d,T )
\end{eqnarray}
 
\subsection*{Supplementary Note 2.: Finite-Temperature Benchmarking and Model Justification}

Finite-temperature $\beta$-decay data are not accessible experimentally, making
theoretical benchmarking essential. While zero-temperature comparisons validate
ground-state properties, temperature-dependent weak rates require validation against
microscopic models that explicitly account for thermal excitation, pairing reduction,
and redistribution of Gamow--Teller (GT) strength.

\subsubsection*{A. Comparison with Finite-Temperature QRPA Calculations}

To assess the reliability of the present approach, we compare our calculated
$\beta$-decay half-lives with finite-temperature QRPA-based studies. For the nuclei
$^{120}$Sr, $^{122}$Zr, and $^{124}$Mo, we consider the FTQRPA results of Minato and
Hagino \cite{MinatoHagino2009}, which are based on a finite-temperature
Skyrme-Hartree-Fock + BCS framework. For $^{130}$Sn, $^{132}$Sn, and $^{138}$Sn,
we compare with the FT-PNRQRPA calculations of Ravli\'{c} \textit{et al.}
\cite{Ravlic2021}, performed at a representative stellar density
$\rho Y_e = 10^7~\mathrm{g\,cm^{-3}}$ using the D3C functional.

\begin{figure*}[h!]
	\centering 
	\includegraphics[width=0.85\textwidth, height=0.6\textheight, keepaspectratio]{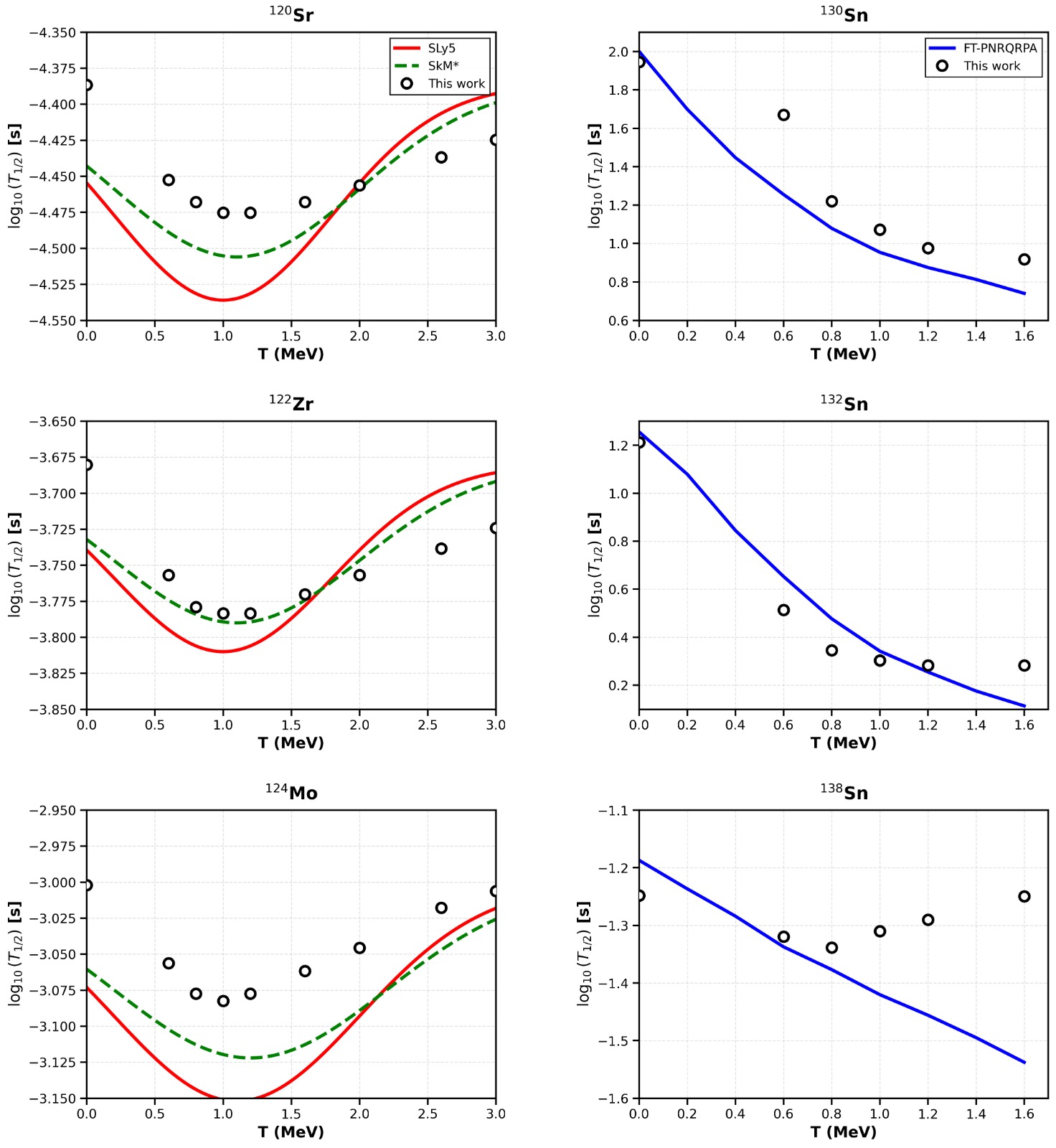}
	\caption*{Supplementary Figure 1: Comparison of calculated $\beta$-decay half-lives as a function of
temperature with finite-temperature QRPA-based results. Open circles denote the
present work. For $^{120}$Sr, $^{122}$Zr, and $^{124}$Mo, solid and dashed curves
represent the FTQRPA calculations of Minato and Hagino \cite{MinatoHagino2009}
based on the finite-temperature Skyrme-Hartree-Fock + BCS framework. For
$^{130}$Sn, $^{132}$Sn, and $^{138}$Sn, the solid lines correspond to the
FT-PNRQRPA results of Ravli\'{c} \textit{et al.} \cite{Ravlic2021} obtained using the
D3C functional at a stellar density of $\rho Y_e = 10^7~\mathrm{g\,cm^{-3}}$.}  
	\label{FigS1}%
\end{figure*}

The comparison (see Supplementary Fig. 1) shows that despite quantitative differences in absolute
lifetimes—arising from different treatments of pairing, GT strength fragmentation,
and particle-number restoration—all approaches predict a systematic decrease of
$\beta$-decay half-lives with increasing temperature. This qualitative agreement
supports the physical consistency of the temperature dependence obtained in the
present work.

\subsubsection*{B. Relation to Shell-Model Calculations}

Finite-temperature shell-model studies predominantly focus on stellar electron-capture
rates rather than $\beta$-decay half-lives, incorporating explicit density and electron
chemical potential effects. Consequently, direct numerical comparison with our
temperature-dependent half-lives is not meaningful. Nevertheless, shell-model Monte
Carlo and large-scale diagonalization studies consistently demonstrate that thermal
population of excited states leads to GT unblocking and enhanced weak-interaction
rates \cite{Langanke2003}. This underlying physical mechanism is common to both shell-model approaches
and the present framework, albeit implemented through different theoretical
approximations.

\subsubsection*{C. Role of Electron Capture and Applicable Regimes}

At higher densities and for heavier nuclei, electron capture dominates weak processes.
However, the nuclei considered in the present work lie in a region where $\beta$-decay
remains relevant at low and moderate densities, particularly during pre-collapse and
early high-temperature evolutionary stage. Our results should therefore be viewed as
complementary to electron-capture studies, providing insight into thermal effects on
$\beta$-decay lifetimes prior to the onset of electron-capture–dominated dynamics.

\subsubsection*{D. Summary}

The agreement in temperature trends between the present approach and microscopic
finite-temperature QRPA calculations, together with qualitative consistency with
finite-temperature shell-model physics, demonstrates that the model captures the
essential thermal effects on $\beta$-decay lifetimes within its intended domain of
applicability.

\bibliographystyle{elsarticle-num}